# Envelope reconstruction of the $_0S_2$ – free oscillation of the earth

*Herbert Weidner*

*Am Stutz 3, D-63864 Glattbach*

*herbertweidner1@freenet.de*

The earthquake in December 2004 caused free oscillations of the Earth. Vibrates the earth in the "football mode" $_0S_2$, splits the natural frequency due to rotation into five closely adjacent individual components. The sum of these spectral components in the frequency band near 309 µHz produces a beat that gives the overall amplitude envelope a characteristic, regular pattern. From the measured envelope the parameters frequency, amplitude, phase and damping of generating sinusoids can be reconstructed. Since the method is extremely sensitive to changes in frequency and phase, these quantities can be determined precisely. The results depend on the geographical location of the site. Further results are the half-life of the amplitude decrease and the resonator Q. It is shown that the interaction of the five individual frequencies can be interpreted as amplitude modulation, which requires a nonlinear process in the Earth's interior.

***Introduction***. The globally distributed superconducting gravimeter[1] of the Global Geodynamics Project (GGP) networks[2] measure the vertical component of the gravitational field with excellent accuracy and very high temporal resolution. Every reading owns a timestamp, which is normally ignored in spite of the fact, that it contains valuable information. The widespread used method of Fourier analysis (FFT) simply can not use timestamps. One way to evaluate this information is to analyze the frequencies of short episodes and encode the amplitudes at any point by a false color. The parallel multi-channel version is called a spectrogram and allows a first overview over the required bandwidth and an estimation of the half-life of the vibrations.

In the spectrogram shown in Figure 1, one can see a regular pattern that is repeated at intervals of (60 ± 0.5) hours. The maximum amplitude of the frequencies 300 µHz, 309 µHz and 318 µHz alternate in a certain temporal order. There a hints that there could be another, weaker pair of frequencies at 305 µHz and 314 µHz. Other European SG sites provide similar spectrograms with a good quality for this event.

The pattern of a weaker earthquake on 2005-03-28 (Figure 2) suggests distance of 30 hours as repetition time. The cause may be the different geographical position of the epicenter with respect to Europe, with dislocated nodal lines of the $_0S_2$ resonance on the earth's surface. It is striking that the pair of frequencies at 305 µHz and 314 µHz is clearly missing.

The recorded data of non-European SG sites as s1, s2, tc, ma, cb, ka show a completely different structure, see Figure 3. The central frequency 309 µHz is only guess. The 60 h rhythm marks the picture. The "cloud pattern" at 290 µHz and 229 µHz is generated largely by strong secondary peaks of the used window function that must be applied prior to each FFT. Other window functions have weaker secondary peaks, but also significantly lower frequency resolution. The widely-used Hamming window is characterized by very weak secondary peaks, but it can also produce strange patterns. Figures 1 and 4 *only* differ in the choice of window function.

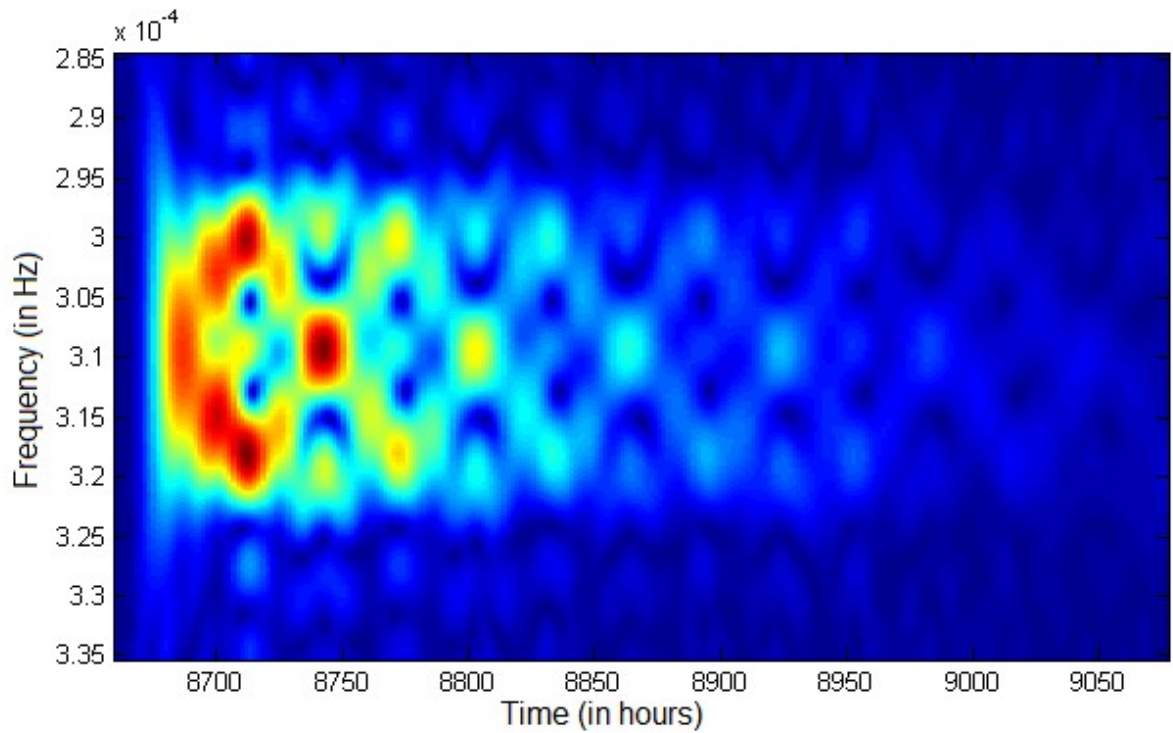

*Figure 1: High-resolution spectrogram of the "football mode" recorded in Strasbourg (st) after the strong earthquake on 2004-12-26. The clock starts on 2004-01-01, the leftmost part is missing due to overloading of the measuring device.*

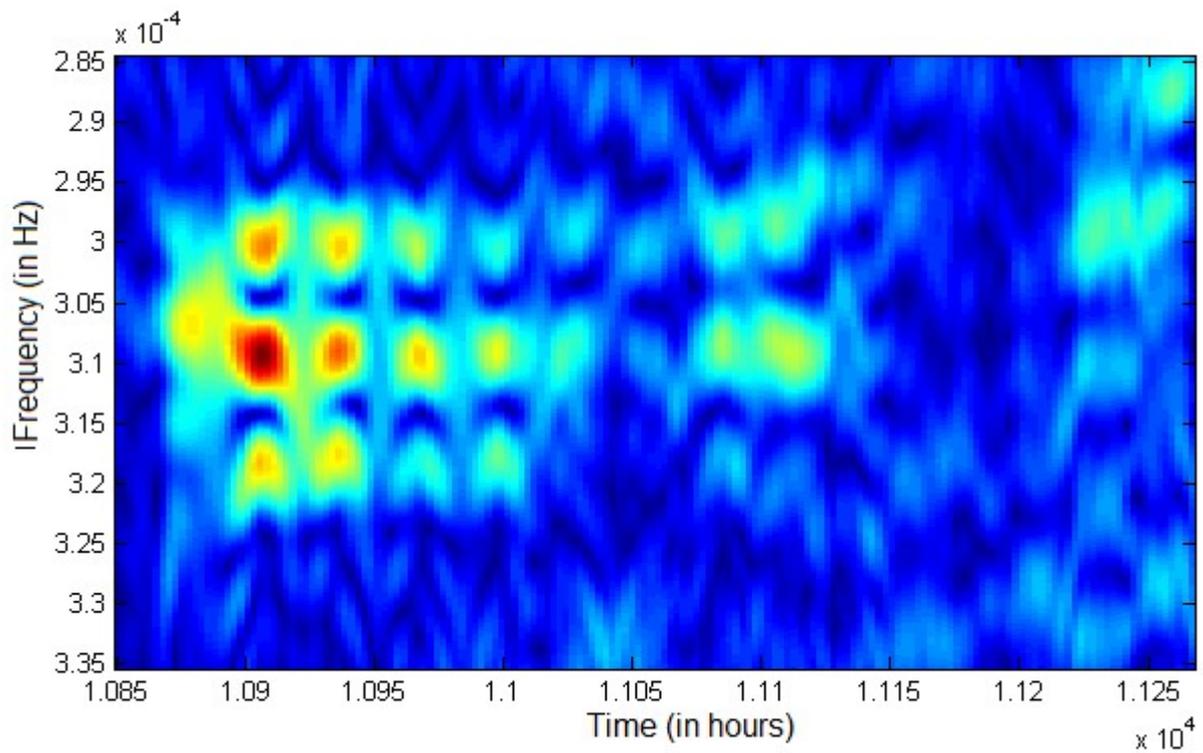

*Figure 2: High-resolution spectrogram taken at Bad Homburg (h2) after the earthquake on 2005-03-28. The gaps after the 11030th hour result of missing data.*

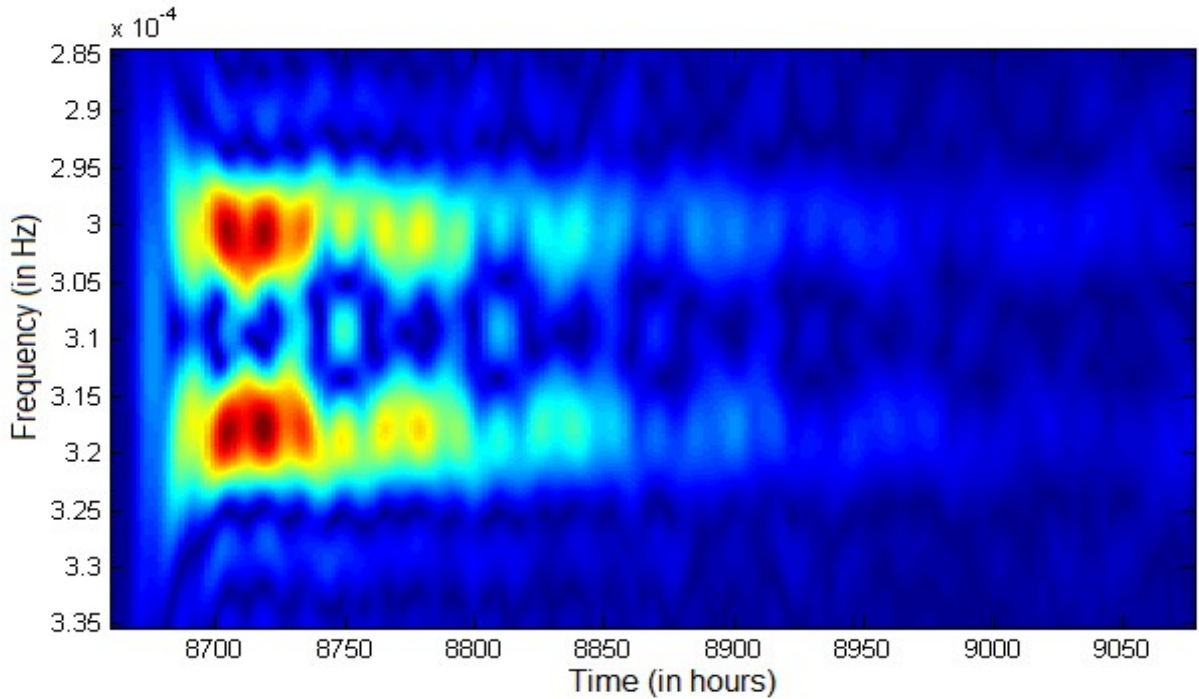

*Figure 3: Spectrogram taken in Canberra (cb). The left-most part has been deleted due to overloading of the device.*

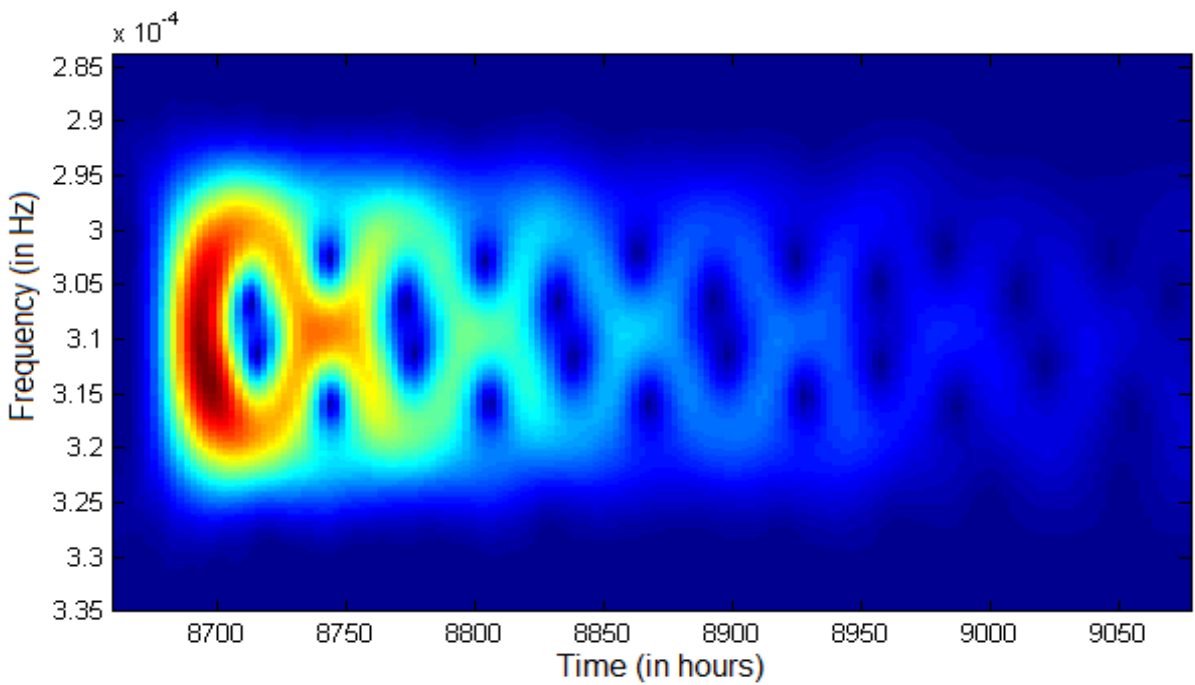

*Figure 4: Spectrogram of the "football mode" recorded in Strasbourg (st) after the strong earthquake on 2004-12-26. The data was hamming-windowed before FFT.*

The shown periodic patterns - regardless of the window function - can be difficult to interpret. Absurd would be the declaration, frequency generators in the Earth's interior would be switched on and off, or change their radiation direction periodically during a few hours.

***Extension of the duration of the episode***: To determine the spectral frequencies the data of each SG-site is divided into episodes of suitable length. Then, the spectral components are determined by Fourier analysis. In the previous pictures the episode lengths were 2800 samples or 47 hours. Thus, short episodes lead to good time resolution and bad frequency resolution. Extending the episodes to 40000 samples with reduced sampling time of 30 s, improves the frequency resolution and allows an accuracy of about 0.3 µHz. This gain is paid for by a substantial blurring on the time scale, because now FFT is calculated over a period of about 300 hours. In this way, as seen in Figure 5, it is possible to discover slow, small changes in frequency: The frequencies 313.9 µHz and 304.6 µHz do not appear to be constant, but are possibly frequency modulated with a period of 60 hours. The frequency changes of both spectral lines are in opposite directions and are estimated to ± 0.3 µHz.

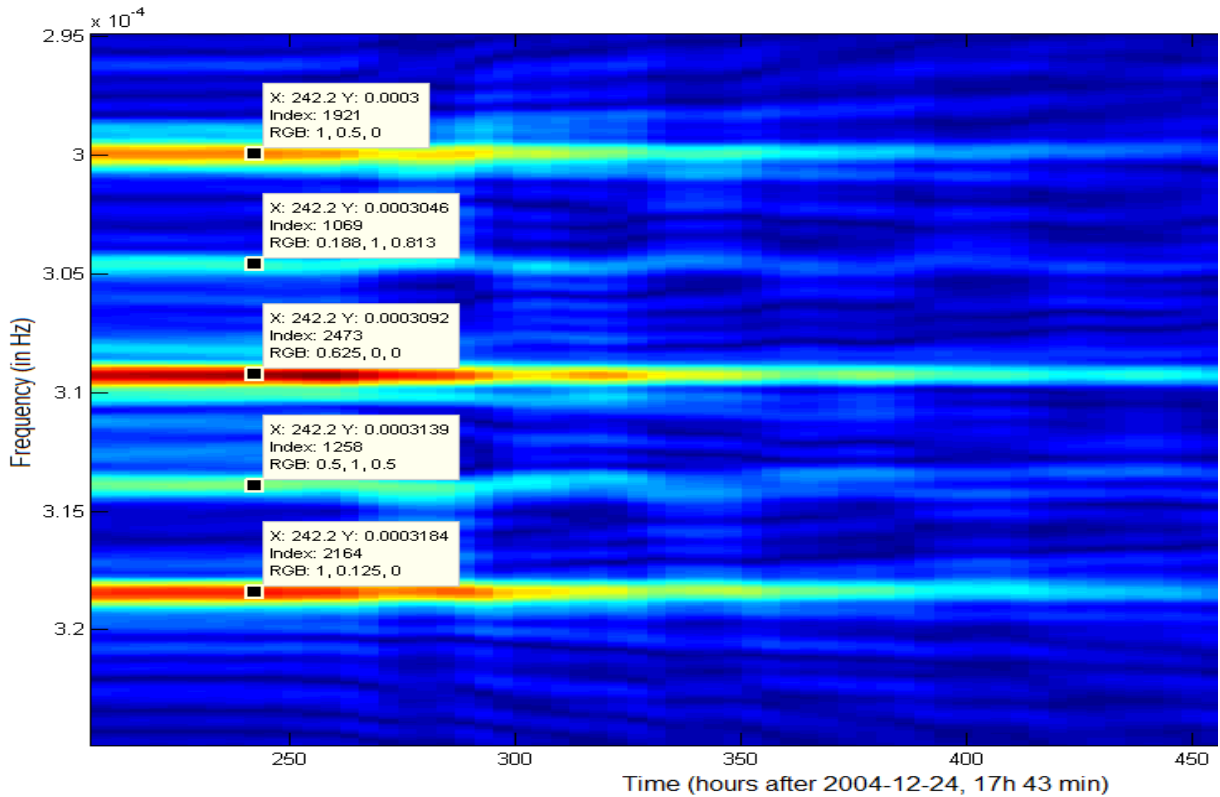

*Figure 5: High-resolution spectrogram recorded in Strasbourg. Sampling time was 30 s. The colour and the index indicate the relative strength.*

The frequency of the spectral lines can not be determined more accurately by extending the episode in order to collect more samples. This would require unmodulated oscillations whose amplitude stay constant for an infinitely long time. But the previous pictures tell another story: They show that the oscillations are no longer detectable after 400 hours and, in the meantime, they demonstrate a 60 or 30 hour rhythm. An accurate frequency determination therefore requires other techniques.

***Reconstructing the Envelope***: Fourier analysis provides average values of the frequencies for an extended period of time and does not allow conclusions about short-term changes in amplitude and phase relationships. The reconstruction of the envelope of a sufficiently long episode, in contrast, enables the simultaneous determination of frequency, amplitude, damping factor and phase of each spectral line. Foundations of this principle were laid by Fourier in the year 1822. Thereafter, any periodic signal can be interpreted as a sum of sine waves of different characteristics. Fourier Synthesis is the base of the following analysis of the recorded data after the excitation by the strong earthquake on 2004-12-26.

Closely spaced frequencies can be distinguished only if sufficient number of oscillations are considered. As in the investigated frequency range near 309 µHz a vibration lasts 54 minutes and an accuracy of better than 0.1% is required, the episode has to cover about 20000 minutes. Interference from unwanted adjacent frequencies can be excluded if the signal bandwidth is limited to about 30 µHz. The digital signal processing provides different filtering programs.

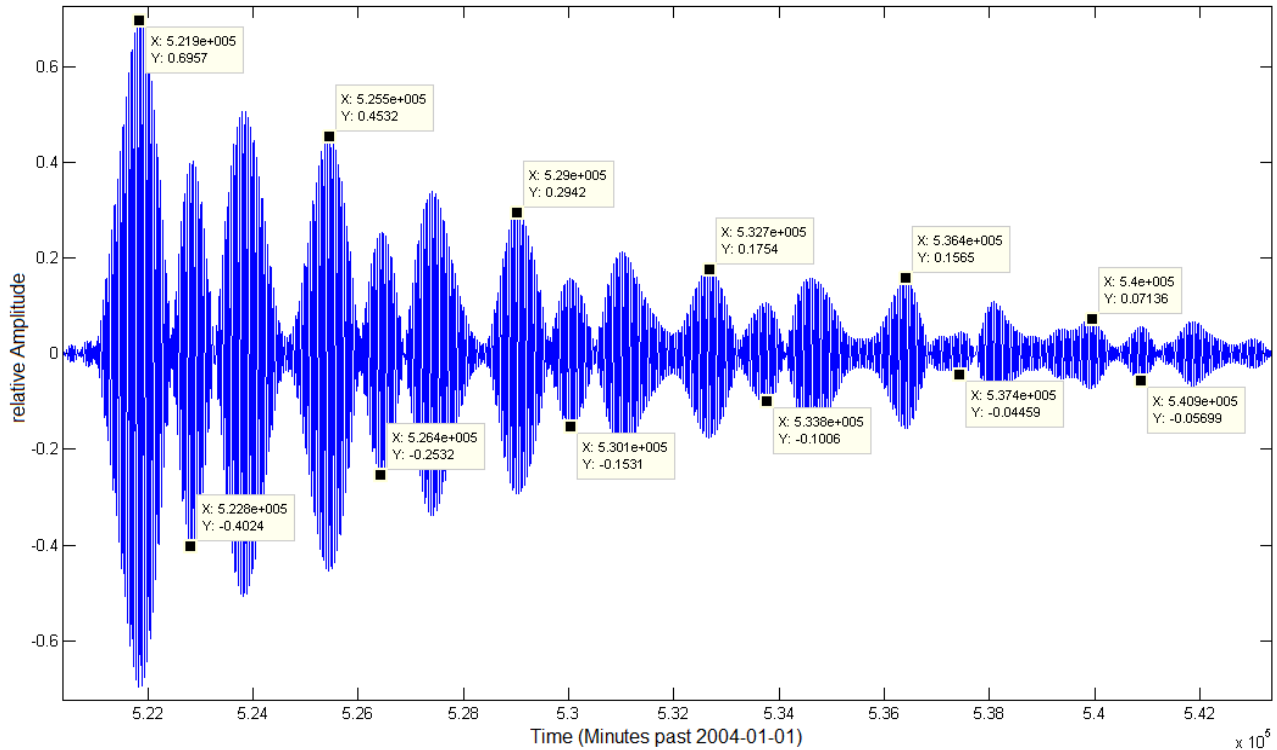

*Figure 6: Seismic data after narrow-band filtering ($f_z$ = 309 µHz; bandwidth = 52 µHz). The data was recorded in Strasbourg after the 2004 Sumatra earthquake.*

The start time of each episode is set individually as soon as the overload of the SG terminates, the data shows no obvious mistakes, and the easily recognizable shape of the envelope is clearly visible. The episode ends when the exponential decrease of the total amplitude is no longer recognizable, or is superposed by interference, indicated by a sudden increase of the amplitude. In Figure 6, an episode registered in Strasbourg is shown. One can see a repeated triple with decreasing amplitude. This form of consistency is only possible if a rigid frequency ratio and a constant phase relationship of the underlying spectral lines during the whole episode is maintained. In particular, the damping factor for all involved vibrations is the same, which simplifies the subsequent calculation. The inserted datatips allow a rough estimation of damping factor and 60-hour rhythm.

A FFT program calculates the spectral components of this episode and plots them in a logarithmic scale (Figure 7). The frequencies and amplitudes of about twenty prominent peaks provide the base for all following calculations. With a smaller number of frequencies it is often impossible to reconstruct the envelope with sufficient precision. Normally, a higher number will not increase accuracy but computing time. First, the frequencies of the peaks must be determined with deviations below 0.4 µHz, because the iteration program does not allow for greater tolerances.

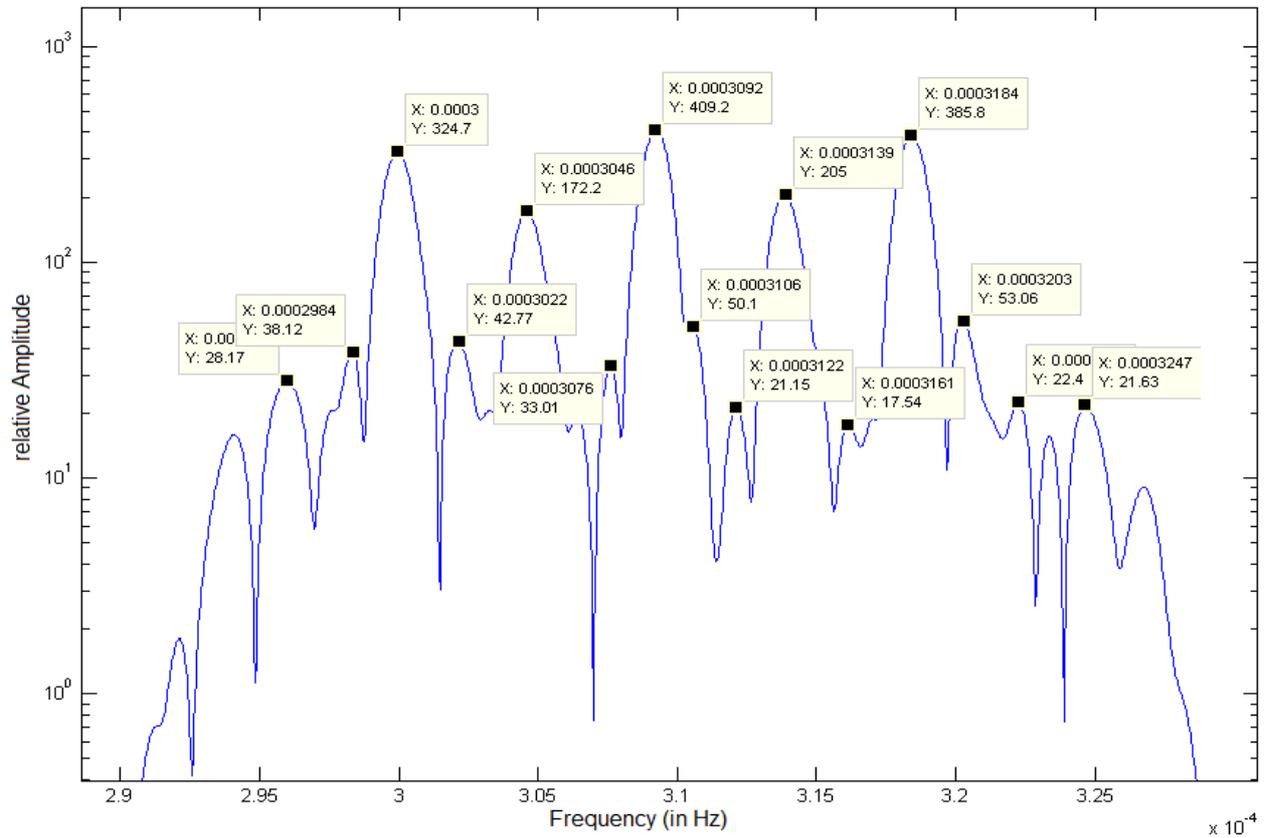

*Figure 7: Spectrum of the oscillation shown in Figure 6.*

We now known the starting values for frequencies and amplitudes, but not the phase angles and the damping factor. These must be selected manually so that the measured envelope of the episode is approximated with moderate accuracy. Only then an iteration can be started, which varies all indicators gradually with the aim of reducing the deviation between measured and calculated envelope.

As can be seen in Figure 6, the envelope is obviously a repetition of self-similar elements with decreasing amplitude. The consequence is a common damping factor for all involved oscillations. Each spectral component follows the approach

$$A(t) = A_n \cdot e^{-\ln(2) \cdot t/T_h} \cdot \sin(2\pi \cdot f_n \cdot t + p_n)$$

The half-time $T_h$ is the same for all involved spectral components. The amplitudes, frequencies and phases may be different. If twenty frequencies are selected, 61 variables must be optimized.

Spectral components that have nothing to do with the natural resonance of the earth, may have different attenuation values, they can also be undamped oscillations (hum). The Iteration program decides, based on built-in criteria, if the measured envelope can be better simulated with different attenuation values for individual frequencies.

After a few hundred iteration cycles the envelope is usually reconstructed very accurate and the program provides the results for frequencies, amplitudes, damping factors and phases of spectral components. In Figure 8, an intermediate result is shown after few iteration runs. One can still see significant differences between measured and calculated envelopes. The iteration can be stopped if no apparent reduction of the discrepancy is achievable.

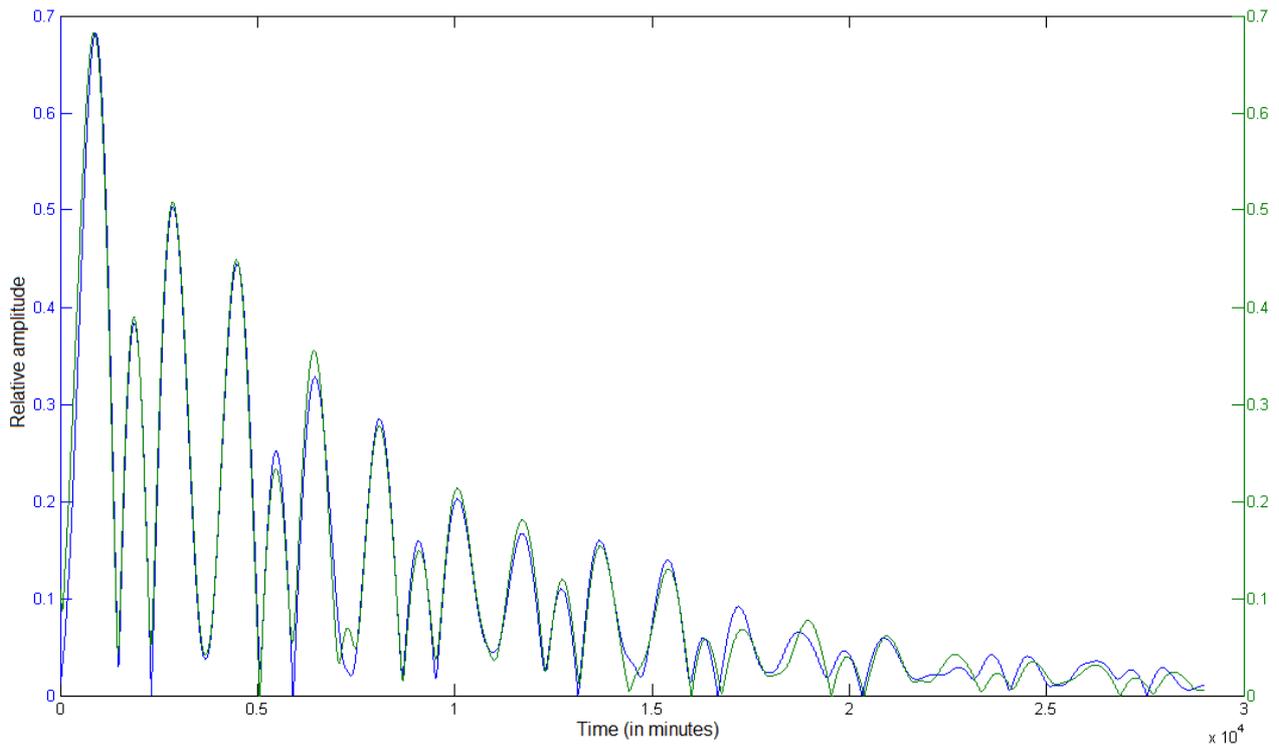

*Figure 8: The measured envelope is coloured blue, the reconstructed envelope is green. The match is still relatively poor after about 20 iterations.*

**Results**: After the earthquake on 2004-12-26, sufficiently low-noise episodes with lengths of about 20000 minutes or more could be found in the records of only 14 SG stations. Those in turn can be divided into two sections:

1. In the spectra of the stations mb, mc, w1, w2, h1, h2, m1, m2, st the center frequency of 309 µHz is clearly detectable and of similar strength as the other four strong spectral lines.

2. In the spectra of the stations s1, s2, cb, ma, ka the center frequency is very weak and can not be clearly identified. The episode at site tc (Chile) is interrupted and therefore too short to be evaluated. The short pieces clearly show the characteristics of section-2.

For every site, about 12 cycles were carried out to suppress systematic errors below the level of numerical noise, each with slightly different boundaries and/or a different number of iterations (min 60, max 800). The individual results of each section were added together, binned with 0.1 µHz steps and are shown in Figure 9.

The striking voids in the immediate vicinity of the four or five strong spectral lines are caused by the shape of each line with typical width of almost 2 µHz, as can be seen in Figure 7. This bell-shape covers weak spectral lines. Because the iteration program allows maximum deviations of ± 0.4 µHz for every spectral line, in the course of the iterations no result can migrate into this area. Between the strong spectral lines, we did not detect any accumulation points except for two small areas at 311.5 µHz (section-2) and 297.9 µHz (section-1). The respective amounts are too small to allow robust conclusions.

Remarkably, the half-widths of the five strongest spectral lines differ significantly. The following discussion is limited just to the properties of these lines. For a more detailed investigation, separate histograms of the step size 0.01 µHz were created, which can be compared pairwise.

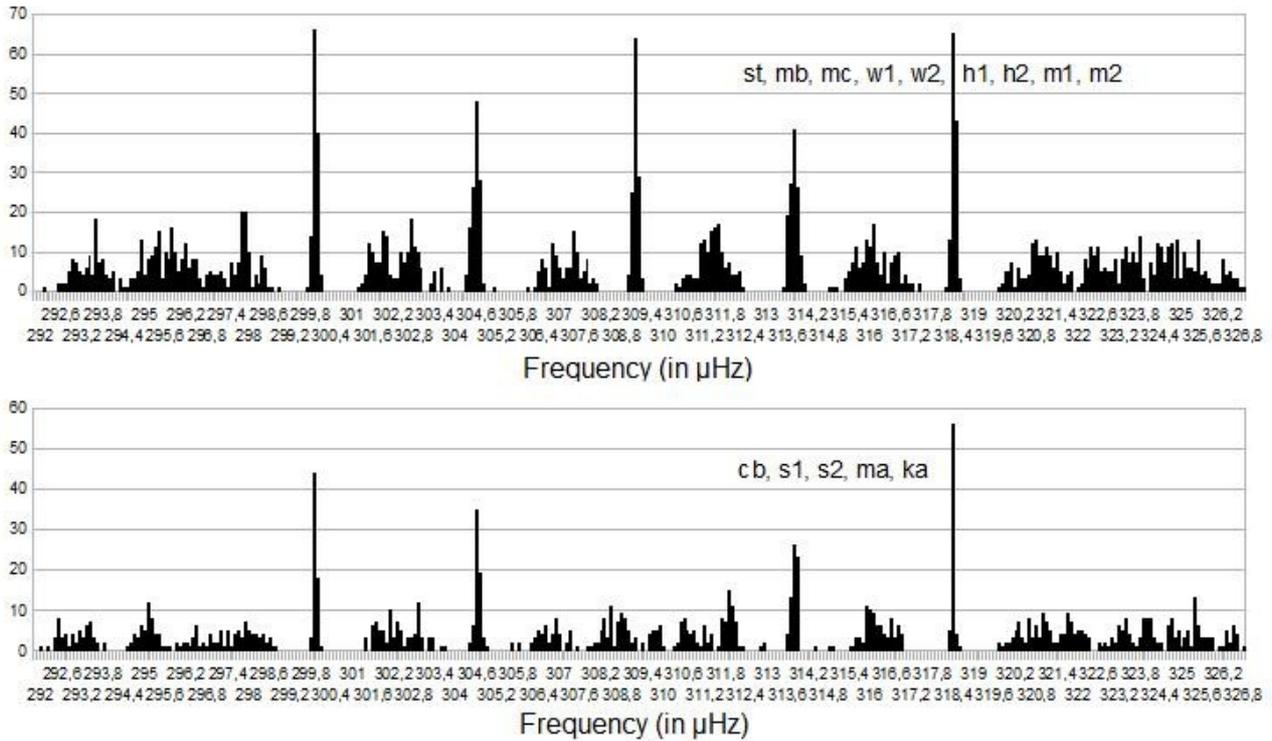

*Figure 9: The upper histogram was calculated from the sum of 177 iteration runs of section-1. The lower histogram (section-2) lacks the central frequency 309 µHz.*

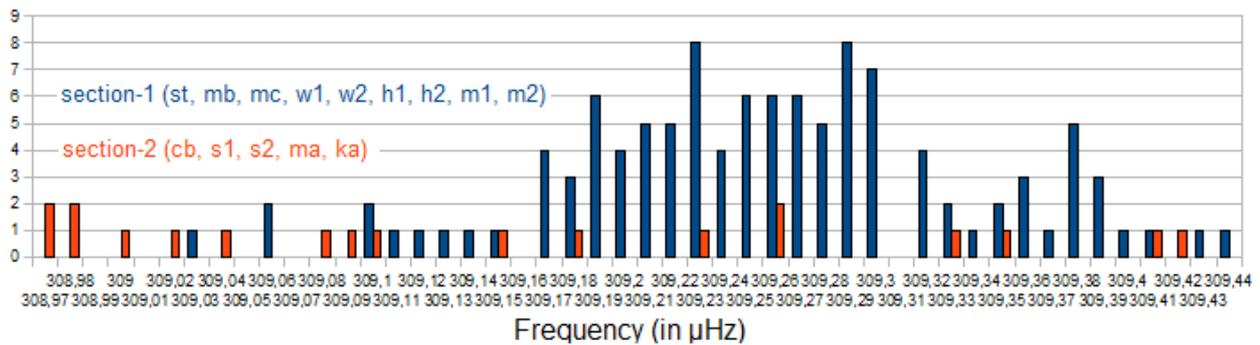

*Figure 10: Histogram around the central frequency 309 µHz. The red-colored bars are the aggregate results of the section-2. The blue-colored bars are the aggregate results of the section-1.*

**The central line 309 µHz**. In Figure 10 you can see striking differences between the results of the stations, located in Central Europe (section-1) and the non-European stations (section-2), in whose spectra the central frequency is very weak. There is no accumulation point and in some spectra of section-2 is even a distinct minimum at 309 µHz, as seen in Figure 11. It does not look like a simple minimum, but rather as a compensation. The mechanism to suppress this spectral line at these sites so extraordinarily well is mysterious. It is unlikely that all four sites are located exactly on node lines of the $_0S_2$-mode with respect of the epicenter.

In section-1, all 112 iteration runs yielded the mean 309.25705 ± 0.08101 µHz, which is listed in Figure 14 as a green dot. The prominent group of between 309.17 µHz and 309.39 µHz includes 106 iteration runs. Their average of 309.2664 ± 0.0581 µHz is shown in Figure 14 as a red dot. These two averages do not differ significantly, the distribution is symmetrical.

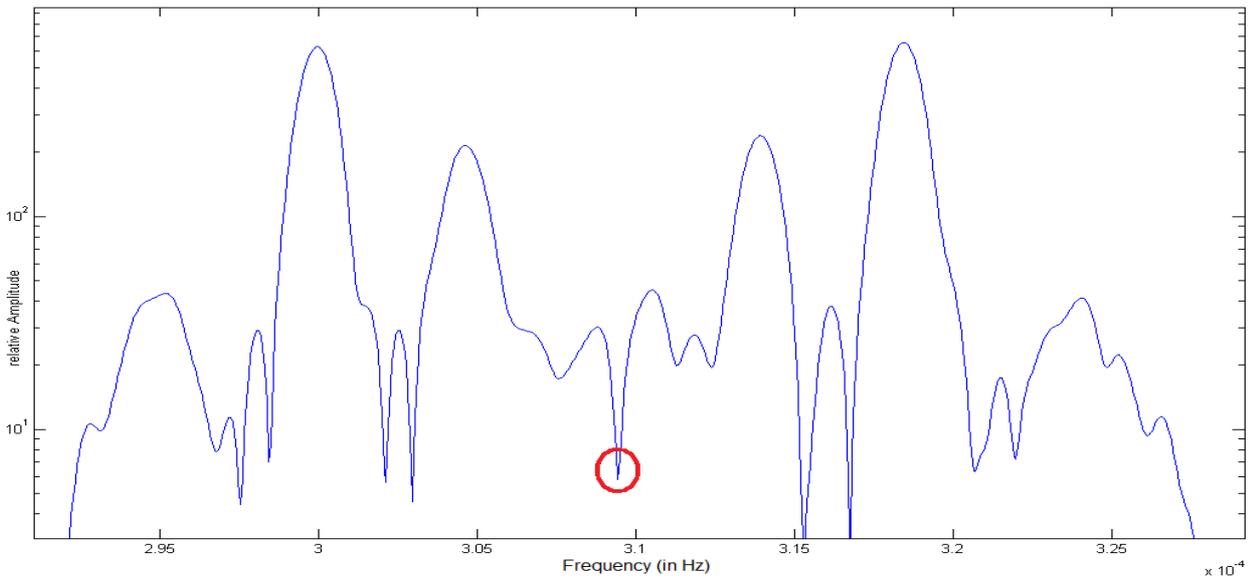

*Figure 11: Spectrum of a 300-hour episode after the earthquake, recorded in Canberra (section-2). The distinct minimum near 309 µHz (red circle) is not a fake!*

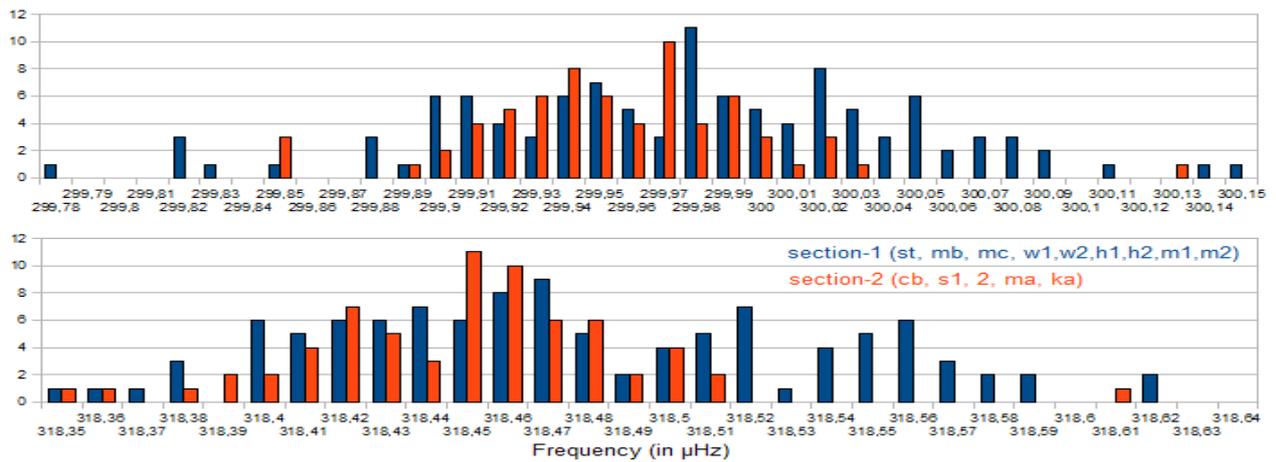

*Figure 12: Histrograms of the "outer" frequencies 300 µHz and 318 µHz.*

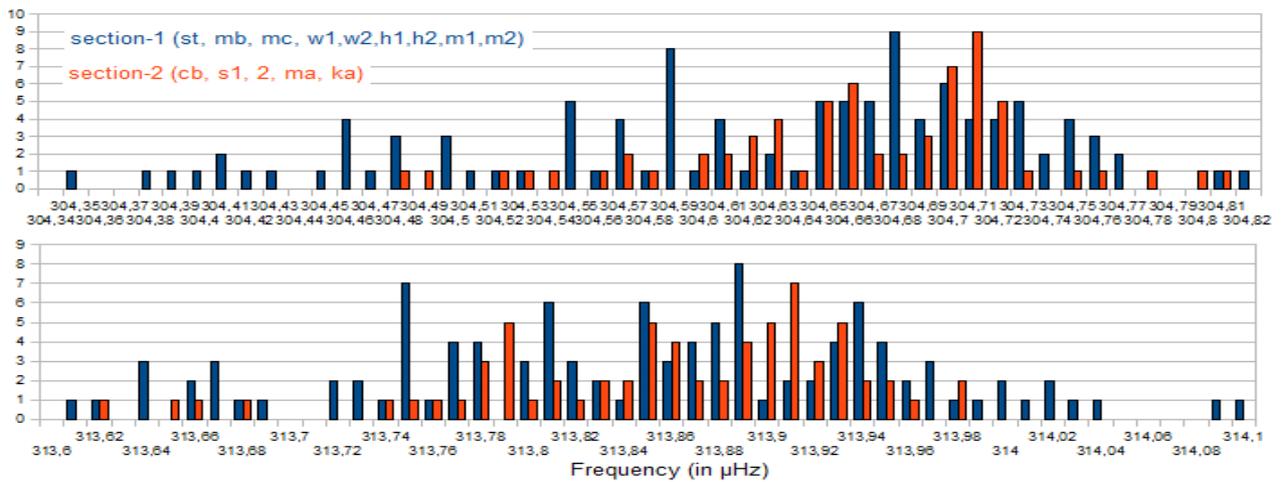

*Figure 13: Histrograms of the "inner" frequencies 305 µHz and 314 µHz*

Comparing the FWHM of the spectral lines, there are remarkable differences. The broadness of the lines 305 µHz and 314 µHz may be a result of a putative frequency modulation (see Figure 5). The error bars for those two frequencies, which were reported by other authors, are also remarkably wide. The blue points plotted in Figure 14 were taken for comparison from[3]: The first three points without error bounds are theoretical predictions. Items 4, 5 and 6 are results of other authors there, including the specified error bounds.

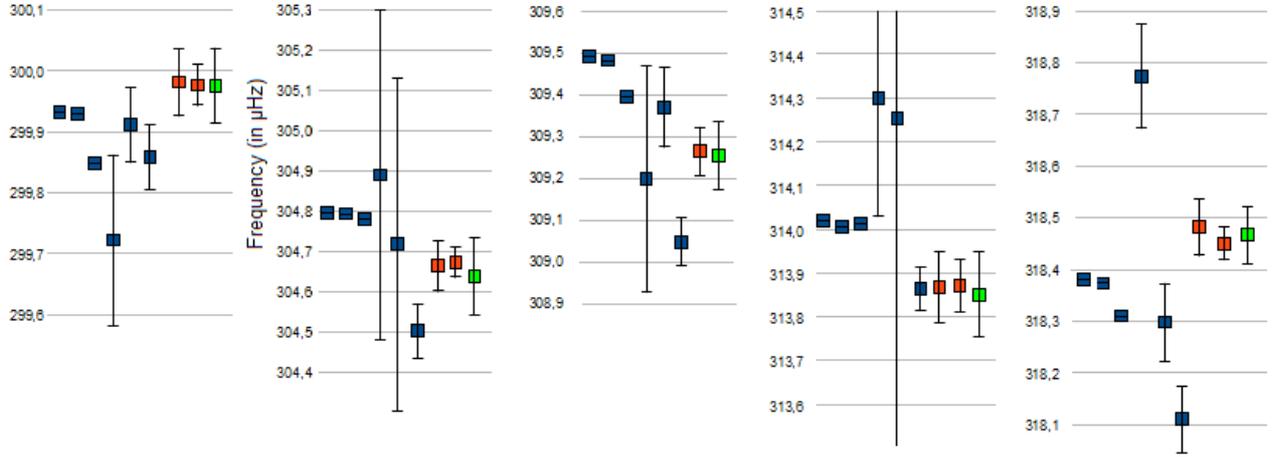

*Figure 14: Theoretical proposals and measured error bars by the following authors: Dahlen & Sailor, Rogister, Rosat et al., Ritzwoller et al., Lindberg, Rosat et al., Weidner*

***Summary of the Results***

| Frequency (µHz) | abs. error (µHz) | rel. error | iteration runs |
|---|---|---|---|
| 299.97592 | 0.06131 | 204 ppm | 179 |
| 304.63819 | 0.09596 | 315 ppm | 177 |
| 309.25705 | 0.08101 | 262 ppm | 112 |
| 313.85119 | 0.09740 | 310 ppm | 177 |
| 318.46686 | 0.05549 | 174 ppm | 175 |

**Drifting frequencies**. The central frequencies change slightly as a function of time. The following values are the averages of the first 333 hours after the earthquake, before the amplitudes were too small for reliable conclusions. In some cases and during short and trouble-free periods, this frequency change can still be proven after 600 hours, although the amplitude decreased under 3% of the initial value. The results also depend on the selected bandwidth (~ 1 µHz), because the adjacent spectral bands influence the result. Figure 5 indicates a possible frequency modulation, therefore this influence was kept low by the choice of an episode length of 60 hours.

The method of measuring such small changes in frequency corresponds to the superheterodyne receiver in high frequency technology: With the help of an extremely narrow band filter with only about 1 µHz bandwidth unwanted spectral frequencies and short-term disturbances are eliminated. From the remaining frequency band of, for example 299.5 µHz to 300.5 µHz, the fixed frequency 290 µHz is subtracted. The oscillation period of the reduced frequency thus increases to about 28 hours, while a possible frequency drift is maintained. At a sampling time of one minute, the distance between two zero crossings should increase to 1670 readings. The actual distances can be easily counted and converted to frequency changes.

| SG-site         | St0405 section-1  | Cb0405 section-2  |
|-----------------|-------------------|-------------------|
| Frequency (µHz) | Drift (pHz/hour)  | Drift (pHz/hour)  |
| 300             | − 21 ± 5          | − 7 ± 6           |
| 304.6           | 43 ± 35           | 78 ± 24           |
| 309.2           | − 2 ± 20          | -----             |
| 313.8           | − 53 ± 22         | − 32 ± 11         |
| 318.4           | 20 ± 5            | 4 ± 1             |

The spectral lines can be classified by their frequency drift in two groups:

- The frequencies 300 µHz, 309.2 µHz and 318.4 µHz are nearly stable and the measurement results depend little on the chosen bandwidth.
- The frequencies 304.6 µHz and 313.4 µHz drift noticeably, they approach slowly. 300 hours after the earthquake, its distance is about 0.03 µHz (100 ppm) lower than immediately after the earthquake. Too small to change the envelope during the first 300 hours .

***Damping and Q factor***. The envelope *E(t)* can be reconstructed very well as the sum of spectral components with the approach

$$E(t) = \sum A_n \cdot e^{-\ln(2) \cdot t / T_h} \cdot \sin(2\pi \cdot f_n \cdot t + p_n)$$

The half-life $T_h$ has similar values at all SG stations that are specified in the table below.

| st    | mc    | w1w2   | mb     | m1m2   | h1h2  | cb    | ka    | ma    | s1s2  | site        |
|-------|-------|--------|--------|--------|-------|-------|-------|-------|-------|-------------|
| 97.82 | 107.2 | 102.15 | 108.42 | 97.53  | 96.99 | 92.32 | 95.87 | 95.62 | 95.92 | $T_h$ (hours) |

The mean of the half-life of one hundred of individual measurements is 98.28 ± 5.60 hours. This represents a Q factor of the Earth in the $_0S_2$-resonance of 502.6 ± 26.5. This result is not compatible with a previously[5][4] estimated value 813 ± 195.

It is noticeable that the half-lives measured by European stations (section-1), without exception, are larger than those measured by non-European sites.

***"Moving" nodal lines***. It was claimed that in the $_0S_2$-mode the nodal lines on the earth's surface migrate west and circle the earth every 60 hours once, corresponding to a speed of 670 km/h [4]. Counter-arguments:

- During the entire episode length of at least 330 hours, the recordings in Vienna were too weak to be evaluated, despite the short distance to Wettzell, where the data were consistently well. Because the distance is only 350 km, a bad (or good) signal quality should arrive after half an hour or less in the neighboring station.
- The Fourier synthesis would fail with a single half-life (about 98 hours) of the damped sinusoidal oscillations. Instead, the duration of an orbit (60 hours) would have to occur somewhere in the formulas and/or in the results.
- It is incomprehensible that there are two fixed sections, whose characteristics do not change during the life of an episode (~400 hours).
- It is incomprehensible that the section-1 is fixed in central Europe.
- It is incomprehensible that the central frequency 309 µHz is *always* registered by the sites of the section-1, but *never* by the sites of the section-2.

- when a node line crosses a SG-site, the phase *must* change significantly, which is not confirmed by the data.

***Amplitude modulation***. A communications engineer, to which the measurement series of SG stations are presented without indication of source, recognizes in the episode after the earthquake on 2004-12-26 in the frequency band around 309 µHz characteristic features of amplitude modulated communications signals and will interpret them as follows:

Section-1 sends the carrier frequency X = 309.257 µHz, AM-modulated with the two frequencies

A = (313.851 – 309.257) µHz = 4.594 µHz and
B = (318.467 – 309.257) µHz = 9.21 µHz.

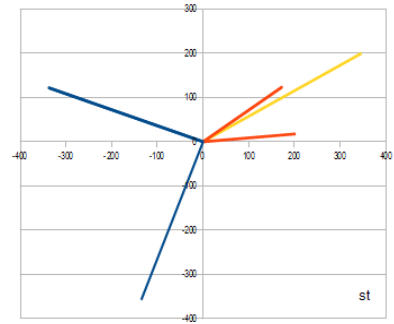

The existence of a carrier frequency and symmetrically located side bands are characteristic features of amplitude modulation (AM). The known phase conditions are fulfilled: the phase angle of the side band frequencies are (almost) symmetrical to the phase of the carrier frequency (309 µHz). The inevitable disruptions hinder detection of an exact symmetry, as shown in Figure 15. The illustrated vector diagram shows the phase relationships during a 200-hour interval (~ 220-fold period). The yellow vector represents amplitude and phase of the central spectral line at 309 µHz. The corresponding vectors for the frequencies 318 µHz and 300 µHz are blue, and those for the frequencies 305 µHz and 313 µHz are red. With decreasing amplitude of the oscillation, the phase can be determined more inaccurate, since the contribution to the shape of the envelope decreases.

*Figure 15: Phase relationship recorded in Strasbourg.*

In the signals of the section-2 lacks the central spectral line at 309 µHz, why it is called "double sideband modulation with suppressed carrier" (DSB) - in the language of communication equipment. The modulation frequencies A and B are identical to those of the section-1. The Figures 16 and 17 show the phase relationships for the same period as in Figure 15, as measured by the stations Canberra and Kamioka. The influence of geography is clearly visible: As the line connecting Strasbourg - Canberra leads almost through the center of the earth, the phases of the signals are effectively "reversed", whereas in Japan there are completely different phase relationships.

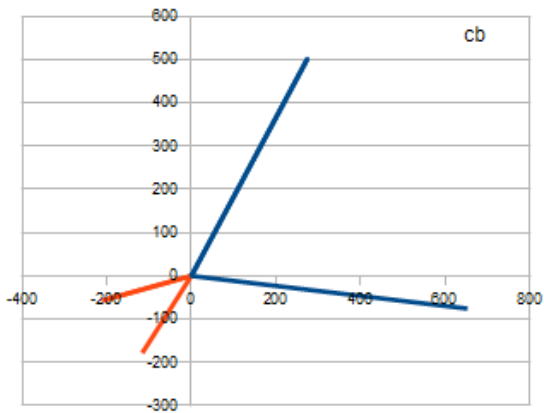

*Illustration 16: Phase relationship recorded in Canberra.*

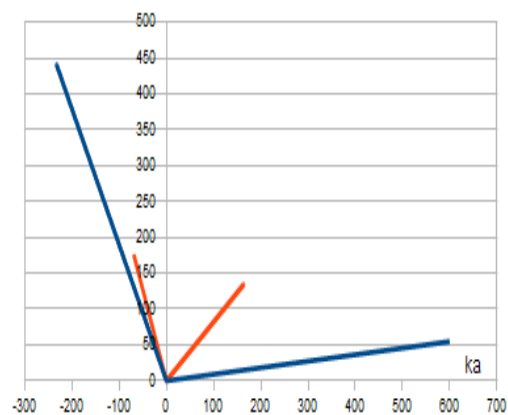

*Figure 17: Phase relationship recorded in Kamioka.*

In AM, the actual message content lies in the quantities A and B, while the carrier frequency is of secondary interest. The reciprocal of A is the apparent rhythm of 60.09 hours in Figure 1, the inverse of B is the rhythm of 30.16 hours in the second image. Measurement errors are the reason that the integer ratio $T_A : T_B = 2:1$ will be missed by 0.4%. In fact, the ratio has to be integer, otherwise the phase angle would change constantly. This change is not observed.

*Numerology*. Remarkably, almost exactly true: B = 2*A. Assuming - without evidence - that the relationship B = 2*A is exact, it is possible to compute "ideal frequencies" through a combination of means and then to compare them with the actual measured frequencies. First, you calculate the differences of consecutive spectral frequencies and the average values of each column.

| Frequency (µHz) | difference (µHz) |
|---|---|
| 299.97592 | |
| 304.63819 | 4.66227 |
| 309.25705 | 4.61886 |
| 313.85119 | 4.59413 |
| 318.46686 | 4.61567 |
| Mean frequency | mean difference |
| 309.23784 | 4.62273 |

Starting from this central frequency, the ideal frequencies can be obtained simply by addition.

| Ideal frequencies (µHz) | measured frequencies (µHz) | absolute error (µHz) | Relative error |
|---|---|---|---|
| 299.99237 | 299.97592 | -0.01645 | -55 ppm |
| 304.61511 | 304.63819 | 0.02308 | 76 ppm |
| 309.23784 | 309.25705 | 0.01921 | 62 ppm |
| 313.86058 | 313.85119 | -0.00939 | -30 ppm |
| 318.48331 | 318.46686 | -0.01645 | -52 ppm |

What a coincidence. The deviations of the measured frequencies of the respective ideal frequencies are remarkably small, and in each case significantly lower than those calculated from standard errors of the measured values. But this is only a strong argument for the correctness of the assumption B = 2*A and no proof. No point of the calculation contains model assumptions about the structure of the earth or rotational splitting. The only basis are analogies to the process of amplitude modulation in communications technology.

*Non-linearities*. Superposition of waves is a linear process, producing no new frequencies. Amplitude modulation, in contrast, is a nonlinear process leading to the formation of new frequencies. The mixing of the carrier frequency 309.2 µHz with the frequencies of A and B generates new frequencies and particular phase conditions. This raises the question: How is this nonlinearity realized in the Earth's interior? Since the sidebands are very strong compared to the amplitude of the carrier frequency, in some cases even stronger, there must be a strong non-linearity.

The observation in Section-2, that the carrier frequency is suppressed perfectly (Fig. 10), leads to a proposal for the non-linear function: it is a parabola. Their quadratic curve generates from the sum of the two frequencies X and A two new frequencies (X + A) and (X − A). In formal language:

$$(\sin(X) + \sin(A))^2 = -½ \cdot \cos(2X) - ½ \cdot \cos(2A) + \cos(X-A) - \cos(X+A)$$

In words: Somewhere in the Earth's interior there are two oscillators:

- After being triggered by a strong earthquake, oscillator X oscillates at the frequency 309.237 µHz. Its amplitude is attenuated with the half-time 98 hours. For some unknown reason, this frequency is only registered by SG sites in section-1.
- Oscillator A oscillates with the frequency of 4.594 µHz, either with constant amplitude or with a half-time which is much higher than 98 hours and therefore can be ignored.

When the sum of both vibrations hits the nonlinear element with a quadratic characteristics, four new frequencies are generated:

1. X + A = 313.86 µHz, phase shift (X, X+A) = +π/2
2. X − A = 304.61 µHz, phase shift (X, X−A) = −π/2
3. 2X = 618.5 µHz. Probably weak and problematic evidence since the spectral line 645 µHz disturbs.
4. 2A = 9.19 µHz. Prove problematic, because there are very strong spectral lines in the immediate vicinity.

The frequency of A is doubled at the same non-linear term:

$$(\sin(A))^2 = \tfrac{1}{2} \cdot (1 - \cos(2A))$$

and modulates the frequency X as well. This creates two additional frequencies:

5. X + 2A = 318.43 µHz
6. X − 2A = 300.05 µHz

Of these, exactly four hit in the narrow area around 309 µHz. If the nonlinearity has a perfect parabolic shape, no further frequencies will be produced. Any deviation from the curve shape would generate more mixed frequencies.

*Remarks*. Open questions are waiting for clarification:
- Why there are two clearly distinguishable sections?
- By what mechanism is the central frequency 309 µHz in section-2 perfectly suppressed and can not be discovered in the noise?
- Why are all the sites of the section-1 in Central Europe?
- The spectrum of the data set from Norway (ny) is unique and shows a totally different structure compared to all the others.

*Acknowledgements*. The author wishes to thank the initiators of the GGP and the operators of monitoring stations for the hard work of data collection. Thanks to ICET in Brussels and GGP-data center in Potsdam for correcting and gathering the files. It was a joy, to evaluate these high quality data.

# Bibliography

1: GWR Instruments, , http://www.gwrinstruments.com/
2: The Global Geodynamics Project, , http://www.eas.slu.edu/GGP/ggphome.html
3: S. Rosat, J. Hinderer, L. Rivera, First observation of 2S1 and study of the splitting of the football mode 0S2 after the June 2001 Peru earthquake of magnitude 8.4, GEOPHYSICAL RESEARCH LETTERS, 2003
5: T. G. Masters, W. Zürn, Free Oszillations: Frequencies and Attentuation., Global Earth Physics: a handbook of physical constants. 1995
4: W. Zürn, R. Widmer-Schnidrig, Globale Eigenschwingungen der Erde, Physikalische Blätter, 2002